# Spontaneous Emergence of Complex Optimal Networks through Evolutionary Adaptation

Venkat Venkatasubramanian[*], Santhoji Katare[◊], Priyan R. Patkar[◊], and Fang-ping Mu

*Laboratory for Intelligent Process Systems, School of Chemical Engineering, Purdue University, West Lafayette, IN 47907, USA.*

## Abstract

An important feature of many complex systems, both natural and artificial, is the structure and organization of their interaction networks with interesting properties. Such networks are found in a variety of applications such as in supply chain networks, computer and communication networks, metabolic networks, foodwebs etc. Here we present a theory of self-organization by evolutionary adaptation in which we show how the structure and organization of a network is related to the survival, or in general the performance, objectives of the system. We propose that a complex system optimizes its network structure in order to maximize its overall survival fitness which is composed of short-term and long-term survival components. These in turn depend on three critical measures of the network, namely, efficiency, robustness and cost, and the environmental selection pressure. Fitness maximization by adaptation leads to the spontaneous emergence of optimal network structures, both power law and non-power law, of various topologies depending on the selection pressure. Using a graph theoretical case study, we show that when efficiency is paramount the "Star" topology emerges and when robustness is important the "Circle" topology is found. When efficiency and robustness requirements are both important to varying degrees, other classes of networks such as the "Hub" emerge. This theory provides a general conceptual framework for integrating survival or performance objectives, environmental or selection pressure, evolutionary adaptation, optimization of performance measures and topological features in a single coherent formalism. Our assumptions and results are consistent with observations across a wide variety of applications. This framework lays the ground work for a novel approach to model, design and analyze complex networks, both natural and artificial, such as metabolic pathways, supply chains and communication networks.

---

[*] Corresponding author, email: venkat@ecn.purdue.edu

[◊] These co-authors made equal contributions to this work

## 1. Introduction: Complex Adaptive Networks

One of the outstanding problems in complex adaptive systems found in engineering, biology, ecology, economics, sociology, and so on, is explaining and predicting the emergence of self-organized network structures with very interesting properties (Albert and Barabási, 2002; Jeong *et al.*, 2000). Such networks are found in a variety of applications such as in supply chain networks (see Fig 1), computer and communication networks, metabolic networks, foodwebs etc., often with similar topological features. Recently, there have been attempts to propose mechanisms for the emergence of the scale-free topologies for such networks. Barabási and Albert (1999) have suggested preferential attachment as a mechanism. Different optimization formulations have been proposed as well by Valverde *et al.* (2001), Carlson and Doyle (2000), and Newman *et al.* (2002). All these results provide valuable insights into the structure of the scale-free networks. However, the questions of why and how the different network configurations emerge, what is the significance of these different topologies, why do we find similar topologies in diverse applications, and what, if any, is the common underlying governing principle remain to be investigated further.

## 2. Theory of Self-Organization of Complex Networks by Evolutionary Adaptation

We propose a general conceptual framework for self-organization of a network by evolutionary adaptation, modelled after Darwin, in which the system's, i.e. the network's (we use these terms interchangeably in this paper), objective is to maximize its chances of overall survival by adapting its configuration according to the environmental pressure. The basic premise is that networks found in nature today exhibit certain characteristic configurations and properties because the same helped them survive the test of time and natural selection. A network typically serves to transport material, energy, and/or information; thus the idea of survival, in all the discussion to follow, is a general one to mean performance towards achieving the design objectives of the network. Therefore, the novel hypothesis is that although human-engineered networks such as supply chains or communication networks have not necessarily 'emerged' by evolutionary adaptation, the underlying design principles that led to their creation could be very similar to those that caused natural networks to evolve to their present forms. The universality of scale-free and other features found in a variety of networks, natural or otherwise, lends support to this view. The proposed framework seeks to shed light on these principles and their guiding influence on network evolution. We will illustrate how the external environment, which imposes or demands certain survival objectives, critically determines the optimal configuration. These insights can be valuable for the study and analysis of *all* networks under various service environments. In this spirit, the framework applies equally to natural as well as human-engineered networks.

We model the overall survival (or performance) as a function of two components, namely, *short-term survival (STS)* and *long-term survival (LTS)*, which are subject to a *cost or resource constraint (C)* in a given survival environment ($\alpha$). This may be formulated as the following optimization problem: Design a system (i.e. a network) by evolutionary adaptation such that it maximizes its survival fitness function, $G$, given by

$$\max \quad G = \alpha\, \eta_{STS} + (1-\alpha)\, \eta_{LTS} - c$$

This conceptual framework is more general than the special cases reported in the literature (Puniyani and Lukose, 2001; Valverde *et al.*, 2001; Amaral *et al.*, 2000). These two objectives, short and long-term survival, are often conflicting, requiring a trade-off in the design. In general, they depend on two critical measures of the system: *efficiency* and *robustness*. We use the term *efficiency* as a measure of the effectiveness of the system's configuration to accomplish its functions. By *robustness* we mean the extent to which the system is able to carry out its functions despite some damage done to it, such as the removal of some of the nodes and/or edges in a network. As noted, these are often conflicting objectives. For example, in the design of engineered systems such as chemical plants, automobiles etc., as one tries to improve the efficiency of the system while keeping the costs down, robustness suffers and vice versa. Design and control engineers are instinctively aware of this compromise. Just as in real-life cost or economic constraints are an unavoidable reality, in nature, too, there are such cost constraints. They show up as constraints in mass, energy, communication links, information capacity, computational power, etc. The efficiency measure essentially determines the short-term survival, while the robustness essentially determines the long-term survival. We summarize all the above ideas in the following postulates of our theory for self-organization of complex systems by evolutionary adaptation:

1. For complex systems such as complex networks, nature adapts the system configuration (i.e. topology) via the processes of evolution and natural selection so as to maximize an overall survival fitness function under a given environmental selection pressure.

2. The overall survival fitness function consists of both short and long-term survival components. These components are dependent on two important measures determined by the configuration – the efficiency and robustness of the system towards performing its functions or objectives. Efficiency governs the short-term survival, whereas robustness impacts the long-term survival.

3. Depending on the system's *functional goal* and its survival *environment*, nature chooses to optimize for efficiency, robustness or both under cost constraints, for "average case" or "worst-case" survival.

The postulates may give the impression that evolution is a teleological or goal-seeking process, but that is not the case. This seemingly goal-seeking behavior of natural systems comes out as an emergent property of the system via evolution when the condition of natural selection is imposed. By "worst case" survival we mean that the network adapts its structure to maximize its chances of survival for a "worst case" failure scenario. Of course, the network can try to maximize its survival chances for an "average case" failure scenario. Many researchers have pointed out how biological scale-free networks are more robust to random failures compared with random networks (Albert and Barabási, 2002; Amaral *et al.*, 2000). This is the "average case" failure scenario. However, it is also known that under intentional or targeted attacks of highly-connected nodes, the scale-free networks break down before the random networks do, for obvious reasons. This is the less frequent, "once-in-a-million-years" kind of "worst case" failure scenario. Designing for protection against "average case" failure scenarios is fine if one expects the biological system's life span to be of the order of thousands of years. However, naturally evolved configurations that exist today have survived geological time scales of hundreds of millions of years, a span during which the "once-in-a-million-years" kind of "worst case" events are quite likely to happen. Therefore the important question that needs to be asked is whether the underlying design criterion might in fact be "worst case" survivability. The premise of this question is that for "routine" biological functions nature may optimize for "average case" failures but for "mission critical" functions it might use the "worst case" scenario as the design criterion. Thus, it is important to recognize that the optimization of efficiency and robustness measures of a system is intimately tied to its *functional goals*. We investigate this trade-off between efficiency and "worst case" robustness in this work.

## 3. Methodology: A Graph Theoretic Formalism

We demonstrate the application of this theory with the aid of a simple, yet illustrative, graph theoretic case study. Consider a system, i.e. a network, consisting of several individual members or nodes. Let us assume that the survival of this hypothetical system depends on the ability of each node to communicate or interact with all other nodes in an efficient and robust manner. The interactions can be through the exchange of material, energy, and/or information. As long as a node is connected to another node, it can communicate with it and pass on messages to others who are part of the overall connectivity. Therefore, the communication between a pair of nodes need not be direct but could occur via one or more intermediate nodes. The set of nodes in direct communication can change, thus the system is adaptive. An example of such a system would be a supply chain, where the interaction of the different nodes via direct links facilitates the transfer of goods for some economic purpose. This adaptive system can be modelled as a graph $S$ of '*n*' vertices and '*e*' edges. The vertices represent the nodes and an edge indicates direct communication between the nodes it connects. Before we proceed any further we would like to lay down the essential concepts and definitions of our graph theoretic framework.

In this paper, we use the terms network and graph, nodes and vertices, interchangeably. Let us now introduce some well-known graph theoretic terms and results (Bollabos, 1985). The *vertex degree* or *degree (k)* of a vertex is the number of edges attached to it. Let us define *n(k)* as the number of vertices with degree $k$ and *p(k)* as the probability that any vertex will have degree $k$. The plot of *n(k)* vs $k$ or *p(k)* vs $k$ is of great interest here. For many real-world networks such as the internet, protein interaction networks etc., this is often found to be a power law distribution i.e. *p(k)* is proportional to $k^{-\gamma}$ with the exponent $\gamma$ of about $2 - 2.5$ (Albert and Barabási, 2002). For random graphs it is well known that this is a Poisson distribution (Bollobas, 1985). The *average path length* or *average vertex-to-vertex distance* of a graph is the average number of steps it takes to go from one vertex to any other vertex. A graph is said to be connected when one can go from any vertex to any other vertex by following the edges. For a graph with *n* vertices, the *minimum* number of edges required to obtain a connected graph is $e_{min} = n\text{-}1$. This reduced set which defines the skeleton of a graph is typically known as the minimum spanning tree *(MST)*. A strongly connected component of a graph is defined as the maximal subgraph in which there is a path between any pair of vertices. Therefore a connected graph has only one strongly connected component which is the graph itself. When all the vertices are directly connected to each other by at least one edge, the graph is said to be a complete graph (*cg*). The number of edges required for a complete graph is given by *n(n-1)/2*. For the sake of simplicity, we make the following assumptions in our problem formulation. Let us assume that all vertices and edges are of equal importance. There are no directions on the edges, and hence the network is not a directed graph. All the following definitions refer to a graph with *n* vertices unless otherwise explicitly specified.

### 3.1. Efficiency

The adjacency matrix of a graph is defined as the binary square matrix:

$A = \{a_{ij}\}, \quad 1 \le i,j \le n, \quad \text{where}$

$$a_{ij} = \begin{cases} 1 & \text{if an edge connects vertices i and j,} \\ 0 & \text{otherwise.} \end{cases}$$

The distance *d(i,j)* between vertices *i* and *j* of a graph is the length of the shortest path between the two vertices. The average path length of a graph is the average distance between any pair of vertices:

$$\langle d \rangle = \text{Average APSP} = \frac{\sum_{i,j} d(i,j)}{n(n-1)}, \quad 1 \le i,j \le n, \quad \text{where}$$

d(i, j) is the distance between vertices i and j.

We define the *absolute efficiency* of a graph as the inverse of its average vertex-vertex distance:

$$Eff = \frac{1}{\langle d \rangle}$$

and the *normalized efficiency* as the ratio of its absolute efficiency to that of a 'Star' network with the same number of vertices:

$$\eta_E = \frac{Eff}{Eff_{star}} = \frac{\langle d \rangle_{star}}{\langle d \rangle}$$

We use the above normalization since the 'Star' configuration, discussed in detail later, is known to give the minimum average path length for an *MST* (i.e. *β = 0*) (Comer, 2001; Deo, 1974).

### 3.2. Robustness

To explain the calculation of robustness we need the following definitions. A vertex *j* is said to be reachable from node *i* iff there exists a path from *i* to *j*. Let $n_i$ be the number of vertices in a strongly connected component $i$ of a graph. Henceforth, we refer to a strongly connected component simply as component. Then ($n_i$-1) vertices are reachable from any vertex in this component. We define this number as the *accessibility* of the connected component $i$: $\mu_i = n_i - 1$

The *effective accessibility* of a graph is then defined as the sum of the accessibilities of its strongly connected components:

$$\hat{\mu} = \sum_{i=1}^{N_C} \mu_i, \quad \text{where}$$

$N_C$ is the total number of strongly connected components.

We define the *structural robustness with respect to vertex j* of a graph as the ratio of the effective accessibility of the graph $S_j$ obtained by deleting vertex $j$ from the original graph to the maximum possible effective accessibility for $S_j$:

$$\rho_j^{st} = \frac{\hat{\mu}(Sj)}{\hat{\mu}_{\max}(S_j)} = \frac{\hat{\mu}(Sj)}{n-2}$$

The *average-case structural robustness* of a graph is defined as the *average* of the structural robustness values computed over all the vertices:

$$\left\langle \rho^{st} \right\rangle = \frac{1}{n} \sum_{j=1}^{n} \rho_j^{st}$$

and the *worst-case structural robustness* as the *minimum* of the structural robustness values computed over all the vertices:

$$\rho_{worst}^{st} = \min_j \rho_j^{st}$$

After removal of a vertex, even if the graph does break into more than one component some or all of these sub-graphs could still be able to perform their respective functions. One may assume that the component with the maximum number of vertices, i.e. the largest component, stands the best chance of retaining its ability to remain functional. Let us denote this largest component of a graph S as $C_{largest}(S)$. The normalized efficiency of such a component could be used as an indicator of the "functional" robustness of the system after damage. We define the *functional robustness with respect to vertex j* of a graph as the normalized efficiency of the largest component of the graph $S_j$ obtained by deleting vertex $j$ from the original graph:

$$\rho_j^{fn} = \eta_E \text{ of } C_{largest}(S_j)$$

Then the *average-case functional robustness* of a graph is defined as the average of the normalized functional robustness values computed over all the vertices:

$$\left\langle \rho^{fn} \right\rangle = \frac{1}{n} \sum_{j=1}^{n} \rho_j^{fn}$$

We further define the *worst-case functional robustness* of a graph as the minimum of the functional robustness values computed over all the vertices:

$$\rho_{worst}^{fn} = \min_j \rho_j^{fn}$$

The general measure of the overall robustness of a graph could be a convex combination of the above-defined four robustness measures. In our work, we have currently considered the overall robustness as simply the worst-case structural robustness:

$$\eta_R = \rho_{worst}^{fn}$$

### 3.3 Redundancy

Since *n-1* is the minimum number of edges needed for a connected graph of *n* vertices, when $e = e_{min} = n\text{-}1$, there is no *redundancy* or *excess connectivity* in the network. We introduce the measure *redundancy coefficient* $\beta$, which is defined as the excess number of edges over an *MST* normalized by the maximum possible departure from an *MST*.

$$\beta = \frac{e - e_{mst}}{e_{cg} - e_{mst}} = \frac{e - (n-1)}{\frac{n(n-1)}{2} - (n-1)} = \frac{2\,(e - n + 1)}{(n-1)(n-2)}, \quad where$$

*cg* stands for a complete graph

Therefore, $0 \leq \beta \leq 1$ so that for a minimum spanning tree $\beta = 0$ whereas for a complete graph, $\beta = 1$. We also introduce the notion of *degree of redundancy*, $R_v$, defined as $R_v = <k>/2$ where $<k>$ is the average vertex degree.

When a redundant edge $e_{ij}$ is added, between vertices $n_i$ and $n_j$, and if a path from $n_i$ to $n_j$ already exists (though not a direct one), we call the new edge as a *functionally* redundant edge. That is, it adds *functional redundancy* to the graph. However, if the new edge is added between two vertices that are already connected directly, then the new edge is said to be *structurally redundant*. This is, for instance, like providing a backup link as an added safety measure. Of course, all this redundancy comes, in general, at extra cost.

### 3.4 Cost

Now we introduce the notion of *cost* or *resource* for the network. In general, both the vertices and edges can have a cost associated with them. However, for a given system of *n* vertices, the number of edges *e* available is the amount of resource available to connect the nodes. Since all edges are equal (an assumption in this work), the total cost is simply *e* (assuming cost/edge = 1). For a minimum spanning tree the resource or cost needed is $e_{min} = n\text{-}1$. Thus, when one uses more than the minimum edges needed to form a connected graph, one would incur an extra cost which is modelled as *C(β,k)*. When *C* is equal to zero, one is providing the system of *n* vertices with the minimal number of edges required to stay connected, which is *n-1* (i.e. *β = 0)*. When *e = n(n-1)/2*, one is providing maximum resources, which is the amount required to make a complete graph. This leads to maximum redundancy, with *β = 1*. Of course, this added redundancy is achieved by incurring the additional cost. Thus, *C* is also a measure of the economy of the design.

### 4. Case Study for Evolutionary Adaptation

As noted earlier, the survival of the hypothetical adaptive system depends on the ability of all its members (nodes) to communicate or interact with all the other members in an efficient and robust manner. Based on the graph theoretical formalism presented earlier, we propose that a measure of the efficiency of the network is the inverse of its average path length. We also define "average case" and "worst case" robustness measures as the *average* and *minimum effective accessibility ratios* arising from a vertex or edge deletion of the original graph. This is essentially a measure of the network's ability to stay connected as a community despite the removal of a node or a communication link between two nodes. Note that this is different from how robustness is defined in the computer networks literature (Comer, 2001) as our context here is different. One can perhaps define other efficiency and robustness measures but these are simple and intuitively appealing for the functional goals of our model system.

Under these conditions, the optimization problem may be formulated as follows: For a given environment $\alpha$ and cost functions $c_1(\beta,k)$ and $c_2(n)$, design a network by evolutionary adaptation that maximizes its survival fitness function $G$ given by:

max  $G = \alpha \ \eta_E \ + \ (1-\alpha) \ \eta_R \ -c_1(\beta,k) \ -c_2(n),$ where

$\alpha$ is a constant, $0 \le \alpha \le 1$

$c_1$ is the cost function related to the addition of edges

$c_2$ is the cost function related to the addition of nodes

$k$ is the vertex degree of the node to which a new edge is being added

$\beta$ is the redundancy coefficient

$n$ is the number of nodes

The optimization variable is the network topology or configuration, i.e. the connectivity or the layout of edges between the nodes. The optimization is carried out subject to the constraint that the configuration must yield a connected network, namely there must exist a path through the edges between every pair of nodes. This is a general formulation of the topology optimization problem. In this work, however, we primarily focus on the non-redundant, minimum spanning tree version (i.e. *β=0*) of this formulation. This simplifies the analysis and makes the interpretation of the results more transparent. The more general, redundant case (i.e. *β≠0*), is under investigation and will be reported at a later date. However, some preliminary results for the redundant case are presented and briefly discussed for $\alpha = 0.5$. In the simplified, i.e. the non-redundant version, one is already provided with a population of $n$ nodes and a set of edges $e \ (=n-1),$ and an environment $\alpha$. That is, the cost or resources available to the network is fixed in terms of $n$ and $e$. Hence, the objective function reduces to

$$\max_{configuration} \quad G = \alpha \ \eta_E + (1-\alpha) \ \eta_R \ \text{for a given } (n, \ e)$$

The $\alpha$ parameter models the environmental or selection pressure on the network. When $\alpha$=1, the survival of the system depends *entirely* on its *efficiency (i.e. short-term survival)* with no regard for robustness. Similarly, at the other extreme, when $\alpha$=0, the survival is determined *entirely* by its "worst case" *robustness (i.e. long-term survival)* with no regard for efficiency. For the intermediate values of $\alpha$, the environment demands that the system be both efficient and robust to varying degrees. Thus, by varying $\alpha$ from 0 to 1, one can impose different selection pressures on the survival of the system and explore the properties of the emergent structures.

This optimization problem could be tackled using various methods. However, since we are interested in the question of how natural networks might have evolved, we investigate this using an evolutionary approach such as the genetic algorithm. Genetic Algorithms (GA) are stochastic search techniques modelled after the Darwinian evolutionary principle of survival of the fittest and have been used for complex optimization problems (Holland, 1992; Venkatasubramanian, et. al., 1993). We conducted GA simulations for two sizes of networks, namely, *n=30* and *n=200*. Even the smaller network (*n=30*) is rich enough to reveal interesting structures as our results demonstrate. The smallness of the model system need not be a limiting factor as numerous studies in cellular automata (Wolfram, 1984) simulations have shown interesting behaviour emerging even in small systems. This is also seen in foodwebs (Dunne *et al.*, 2002) which have smaller network sizes ranging between 20 and 160 nodes. Furthermore, the smallness works to our advantage as topological features that might be obscured in larger networks are more easily identifiable here. Thus, we use the smaller network to explore emergence of interesting topological features which is obscured in the larger network of 200 nodes. We use the larger network to demonstrate the emergence of different classes of networks which is not apparent for the smaller networks.

In our GA simulation, a network candidate configuration for *n* nodes and *e* edges was represented as an adjacency list. An adjacency list, unlike an adjacency matrix, stores only the list of adjacent vertices, i.e. vertices connected by a direct edge, and is therefore a more compact representation than the adjacency matrix. The GA population had 50 connected network candidates, randomly created at the start of each run. The process of evolution was simulated using mutation and crossover operators suitably designed to operate on the adjacency lists. Mutation of an adjacency list was carried out by removing an existing randomly selected edge and moving it between two previously non-adjacent vertices. Crossover occurred between two networks via the exchange of sub-graph components. Feasibility was maintained by discarding any configuration, created during evolution, which corresponded to an unconnected graph. The population was allowed to evolve for 300 generations for the smaller network (*n=30*) and for 2000 (*n=200*) generations for the bigger one.

We ran the GA simulations for different $\alpha$ values, and found the emergence of several structures, both power law and non-power law. We present the most interesting structures that emerged and discuss their significance. The results are as shown in Fig. 1 and Table 1. We see that when $\alpha=1$, the genetic algorithm discovers the "Star" topology (Fig. 1a) as the most optimal configuration to maximize efficiency. This is what one would expect as it is well-known that the star configuration is the most efficient with minimum average path length at minimum cost (i.e. $\beta = 0$) (Comer, 2001; Deo, 1974). However, the 'Star' has the lowest 'worst case' robustness: if the central node is taken out the whole system collapses. The 'Star' is a scale-free network with a power law exponent $\gamma=1$. At the other extreme, for $\alpha=0$, we get the 'Line' topology (Fig. 1b) for $e$=29 edges, and a "Circle" topology (Fig. 1c) for $e$=30 as shown. The Circle has the maximum 'worst case' robustness: the deletion of any one node does not affect the system's ability to allow all the remaining nodes to communicate with each other. However, the Circle's efficiency is poor. The Line is quite similar to the Circle in its overall properties, second only to the Circle in its robustness but poorer in efficiency. It is important to note that the Line and Circle structures do not fit the standard power law model. It is easy to see that for the Circle the *ln p(k) vs ln k* plot yields just a point and for the Line a straight line of *positive* slope approaching infinity for large values of *n*.

It is important to draw the distinction between robustness and redundancy as explained above. In our view, they are related but represent different features of the network even though in the literature these terms have been used interchangeably. Consider the Star and the Circle topologies. For large *n* (i.e. $n \rightarrow \infty$), they both have essentially the same number of edges. Hence, the redundancy $\beta$ is *zero* for both structures. However, the Star is the *least* robust *topologically* and the Circle is the *most* robust *topologically*. We call this property as the *intrinsic* or *topological* robustness. The Star can be made more robust by adding more edges to it, that is by increasing the redundancy or $\beta$, but this will incur additional cost due to all the extra edges added. Thus, the overall robustness of a network depends on two features: its topological robustness and redundancy.

For $\alpha=0.8$ and 0.7, an interesting topology emerged, with the qualitative features as shown in Figs. 1d and 1e. We call this topology the 'Hub'. Fig.1d shows a Triangular Hub while in Fig. 1e we see a Pentagonal Hub as the fittest structures for these values of $\alpha$, respectively. The Hub seems to balance the need for efficiency and robustness quite well, which is demanded by the environment when $\alpha=0.8$ and 0.7. We would like to draw the reader's attention to the most important feature of the Hub. The Hub is a combination of the Star and the Circle, with the Circle inside (i.e. the triangular or the pentagonal part) and the Star's spokes outside. The Hub is a power law topology with a $\gamma=0.916$ (Triangular) and $\gamma=0.827$ (Pentagonal). We predict that other Hub formations would emerge for other intermediate values of $\alpha$, including the 'Perfect Hub', which has the *same* number of nodes on the Circular part as it does on the Star's spokes. The Perfect Hub has a *uniform* degree distribution, which can be viewed as a power law topology with $\gamma \rightarrow 0$. The perfect hub for *n*=30 is shown in Fig. 1f.

Thus, in the space of possible topologies, the Star and Circle topologies seem to define two extreme classes of structures. The Star topology introduces the features of hierarchy and centralization of functions while the Circle topology introduces decentralization and distribution of functions. The other classes, which are a combination of the Star and Circle features, emerge for intermediate values of $\alpha$ optimizing the need for both hierarchical and distributed organization of functions.

For the larger, *n=200* network, we explored the nature of the *p(k)* vs *k* distribution for different amount of edges (i.e. different values of '*e*') for a fixed value of $\alpha$ =0.5. The results (averaged over 10 runs) are shown in Fig. 3a, b and c. For smaller values of $e$ (i.e. $e$=200) it appears that the distribution could be seen as exponential or power-law in character based on the value of the correlation coefficient, $r^2$. For intermediate values of '$e$' (i.e. $e$ = 220-250), the distribution appears to be a power law with a cut-off. For larger values of '$e$', the distribution approaches the Poisson distribution. We are currently investigating these interesting trends and possible underlying explanations.

## 5. Discussion

We observe that the qualitative features of the topologies that emerged in our GA simulations are found in many applications. For instance, the Star topology is commonly found in computer networks that are often designed for high efficiency and robustness to random failures. The Star is also well-known for its vulnerability to attack or failure at its central node, i.e. the "worst case" scenario. Regarding the Triangular, Pentagonal or the Perfect Hub topologies, we could not find any instances of this formation in the biological networks literature. However, it is interesting to note that this Hub formation is quite prevalent in human engineered systems such as in the layout of airports, shopping malls, and other such large complex structures. For all the other intermediate values of alpha, and for much larger networks, which we have not considered due to computational resource limitations, we expect other $\gamma$ classes to emerge with features that are a mixture of the Star and Circle features. In practice, these features, of course, will be further complicated by domain specific constraints, such as metabolic constraints, cost constraints, and evolutionary quirks like the 'QWERTY' phenomenon. To fully understand all this, a much larger network and more extensive GA simulations are needed.

It is instructive to explore the potential implications of the Line and Circle topologies as the optimal structures for maximum "worst case" survival fitness. This naturally leads to the important question of why and where one might have such an extreme requirement of "worst case" survivability. In engineering applications, such a demand is often made on modules that perform "mission critical" functions. In biological systems, one can argue that the modules that are "mission critical", i.e. be absolutely essential to the long-term survival of the organism, are those that store the genetic code, such as the plasmids, RNA and DNA molecules. It is quite interesting to note that the plasmid has a circular topology, the RNA and the single-stranded DNA have linear topologies, and the double stranded DNA structure can be thought of as a redundant linear topology. While this observation is speculative in nature, it is, nevertheless, interesting that such a possibility would emerge from the robustness analysis.

Our GA simulations also demonstrate that nature could have discovered these structures through evolutionary self-organization by using the Darwinian principle of natural selection. The theory also lays down a framework for relating a system's survival to performance measures such as efficiency, robustness and cost as well as relating these in turn to the topological features of the network and the environment. For instance, it appears that from the $\gamma$ values of the networks one can learn about the environmental pressure on them. Thus, this theory seems to be able to suggest when and why the different network structures might emerge.

The ubiquitous presence of star, hub, and circle topologies in diverse applications seems to suggest an underlying universality concept. That is, these topological features are domain or context independent, independent of the details of the domain specific mechanisms, but are governed by some unifying organization principle.

The theory we have proposed is a general one applicable to a wide variety of self-organizing complex systems. It is not just limited to the evolution of complex networks that are under this tension between the two competing objectives of efficiency and robustness. The theory is equally applicable to other adaptive complex systems where the competing objectives could be some other relevant pair, such as competition vs cooperation, which would determine the emergent properties of business organizations and other such economic systems. An analysis of a similar interplay between order and disorder, energy vs entropic influence, might lead to emergent topologies seen in nature such as in snowflakes. Applying this theory to the compromise between distributing political power to citizens in a democracy versus centralizing the power by the government can possibly explain the emergence of governing structures such as the different branches of the government, state-level agencies, and so on for optimal governance.

When one tries to maximize the objective or the *value* function we have proposed, one is attempting to balance these competing objectives, minimize the overall conflict and maximize harmony. The generality of this formulation and its implications seem to suggest a universal principle for the optimal self-organization of complex systems. We propose that this principle may be called the *Principle of Maximum Harmony* and the objective function the *Harmony Function*.

In conclusion, we have proposed a conceptual framework for modelling and analyzing the emergence of optimal structures found in complex adaptive systems through evolution. This general optimization framework clarifies and relates several conceptual issues that have been treated as special cases in the literature. We introduce a generalized objective function using the concepts of *short-term* and *long-term survival* and relate them to the system's performance measures of efficiency, robustness, cost as well as the system's functional goal and the environment. We discuss the important distinction between designing for average case failures of 'routine' functions versus designing for 'worst case' scenarios of 'mission critical' functions. The distinction between topological robustness and robustness due to redundancy is also clarified. We also demonstrate that optimal structures could be discovered through an evolutionary, self-organized, process. We have shown how a network's performance measures and environmental selection pressures are linked to its topological features. This framework lays the ground work for a novel approach to model, design and analyze human-engineered complex networks such as supply chains and communication networks.

## References

Achenie, L. E. K., Gani, R. & Venkatasubramanian, V. (2002). *Computer Aided Molecular Design: Theory and Practice*. London: Elsevier.

Albert, R. & Barabási, A. L. (2002). Statistical mechanics of complex networks. *Rev. Mod. Phys.* **74**, 47.

Amaral, L. A. N., Scala, A., Barthelemy, M. and Stanley, H. E. (2000). Classes of small-world networks. *Proc. Nat. Acad. Sci.* USA. **97**, 11149.

Barabási, A. L. and Albert, R. (1999). Emergence of scaling in complex networks. *Science* **286**, 509.

Bollobas, B. (1985). *Random Graphs*. London: Academic.

Carlson, K. and Doyle, J. (2000). Highly optimized tolerance: robustness and design in complex systems. *Phys. Rev. Lett.* **84**, 2529.

Comer, D. E. (2001). *Computer Networks and Internets: With Internet Applications*. New Jersey: Prentice Hall.

Deo, N. (1974). *Graph Theory with Applications to Engineering and Computer Science*, Prentice-Hall Series in Automatic Computation, Prentice Hall, New Jersey.

Dunne, J. A., Williams, R. J. & Martinez, N. D. (2002). Food-web structure and network theory: The role of connectance and size. *Proc. Nat. Acad. Sci. USA.* **99**, 12917.

Edgar, T. F., Himmelblau, D. M. & Lasdon, L. S. (2001). *Optimization of Chemical Processes*, 2nd Edition., McGraw Hill.

Holland, J. H. (1992). *Adaptation in Natural and Artificial Systems: An Introductory Analysis with Applications to Biology, Control and Artificial Intelligence*. MIT Press.

Jeong, H., Tombor, B., Albert, R., Oltvai, Z. and Barabási, A. L. (2000). The large-scale organization of metabolic networks. *Nature* **407**, 651.

Newman, M. E. J., Girvan, M. and Farmer, J. D. (2002). Optimal design, robustness, and risk aversion. *Phys. Rev. Lett.* **89**, 2, 028301-1.

Puniyani, A. R. & Lukose, R. M. (2001). Growing random networks under constraints. *cond-mat.* 0107391.

Swaminathan, J. M., Smith, S. F. and Sadeh, N. M. (1996). A multi-agent framework for modeling supply chain dynamics. *Proceedings of the Artificial Intelligence and Manufacturing Research Planning Workshop*, June 24-26, Albuquerque, New Mexico, 210.

Valverde, S., Ferrer Cancho, R. & Solé, R. V. (2001) Scale-free networks from optimal design. *Europhys. Lett.* **60**, 512.

V. Venkatasubramanian, K. Chan, J. M. Caruthers, "Computer-aided  Molecular Design Using Genetic Algorithms", *Computers and Chemical Engineering,*  18(9), 833-844, 1994.

Wolfram, S. (1984). Universality and complexity in cellular automata. *Physica D* **10**, 1.

**Figure 1: Supply-chain network** (Swaminathan *et al.*, 1996)

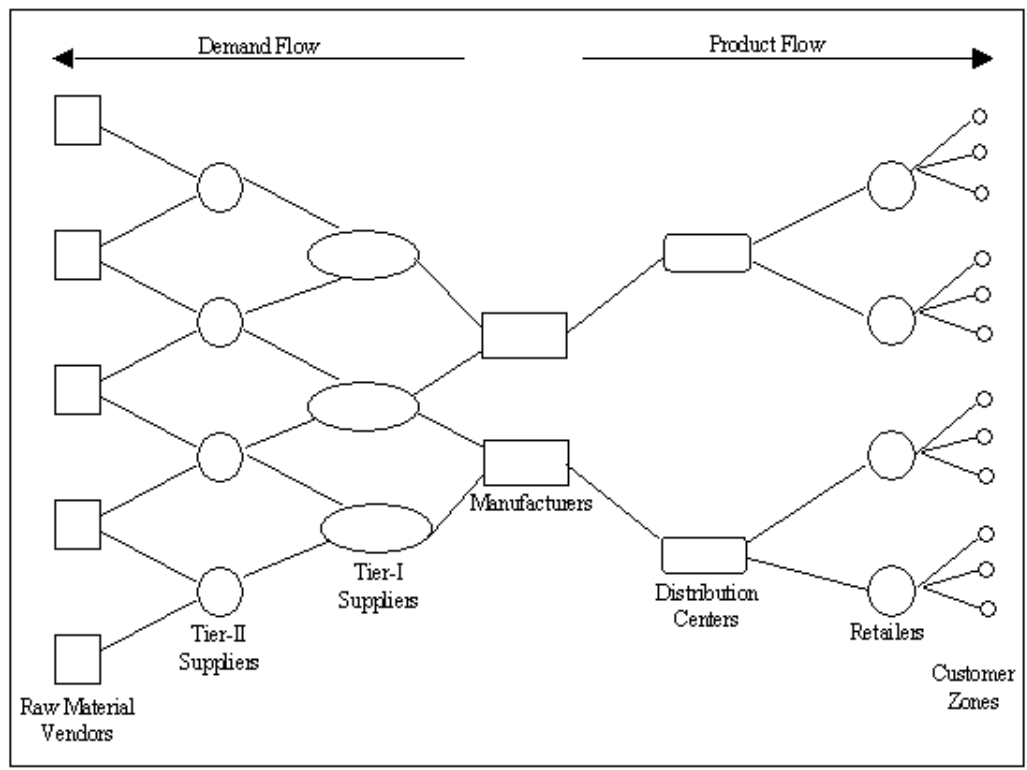

**Figure 2: Network Topologies (for n=30)**

(a) Star (b) Line  (c) Circle (d) Triangular Hub  (e) Pentagonal Hub  (f) Perfect Hub

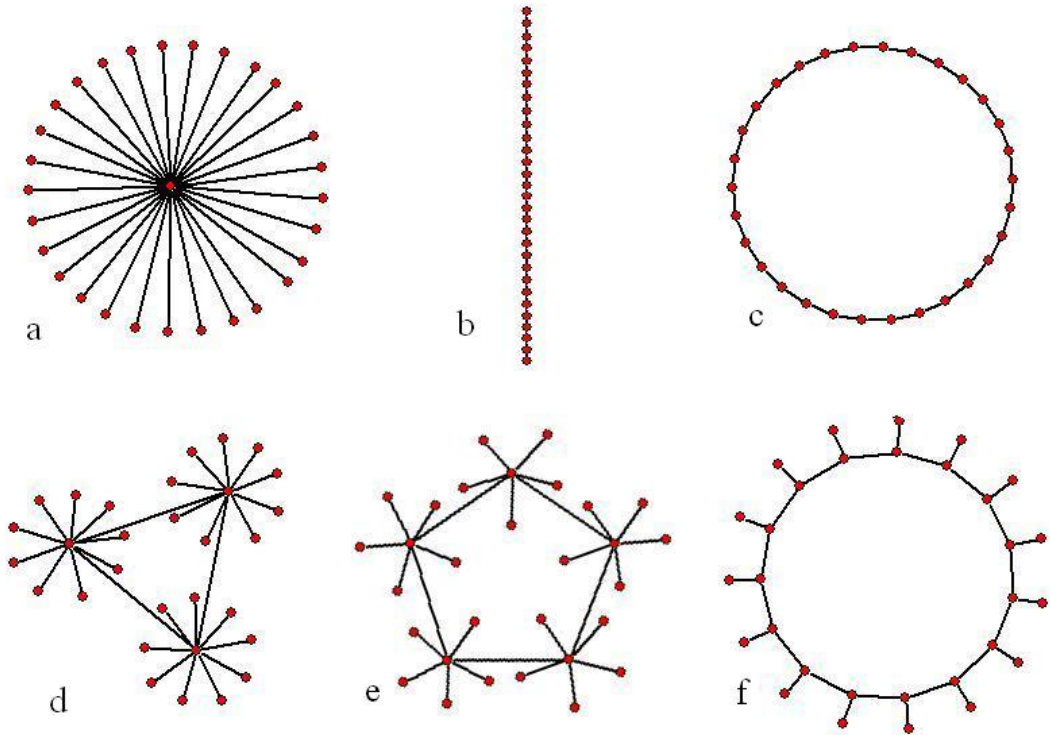

**Table 1: Key Properties of the Networks (for n=30)**

| Structure | α | Efficiency | Robustness | | γ | <k> | Average Path Length |
|---|---|---|---|---|---|---|---|
| | | | Worst Case | Average Case | | | |
| 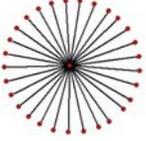 Star | 1 | 1.00 | 0 | 0.97 | -1.0 | 1.93 | 1.93 |
| 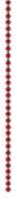 Line | 0 | 0.19 | 0.96 | 0.97 | 3.81 | 1.93 | 10.33 |
| 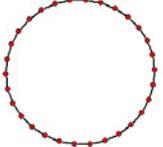 Circle | 0 | 0.25 | 1.00 | 1.00 | - | 2.00 | 7.76 |
| 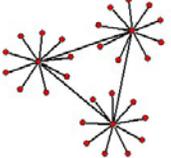 Triangular-Hub | 0.8 | 0.78 | 0.68 | 0.97 | -0.92 | 2.00 | 2.49 |
| 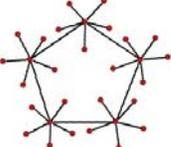 | 0.7 | 0.66 | 0.82 | 0.97 | -0.83 | 2.00 | 2.91 |

| Pentagonal Hub | | | | | | | |
|---|---|---|---|---|---|---|---|
| 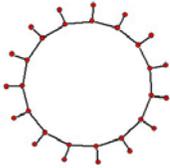 Perfect Hub | ? | 0.40 | 0.96 | 0.98 | 0 | 2.00 | 4.86 |

**Figure 3. The variation of the p(k) vs k distribution with the number of edges (for n=200)**

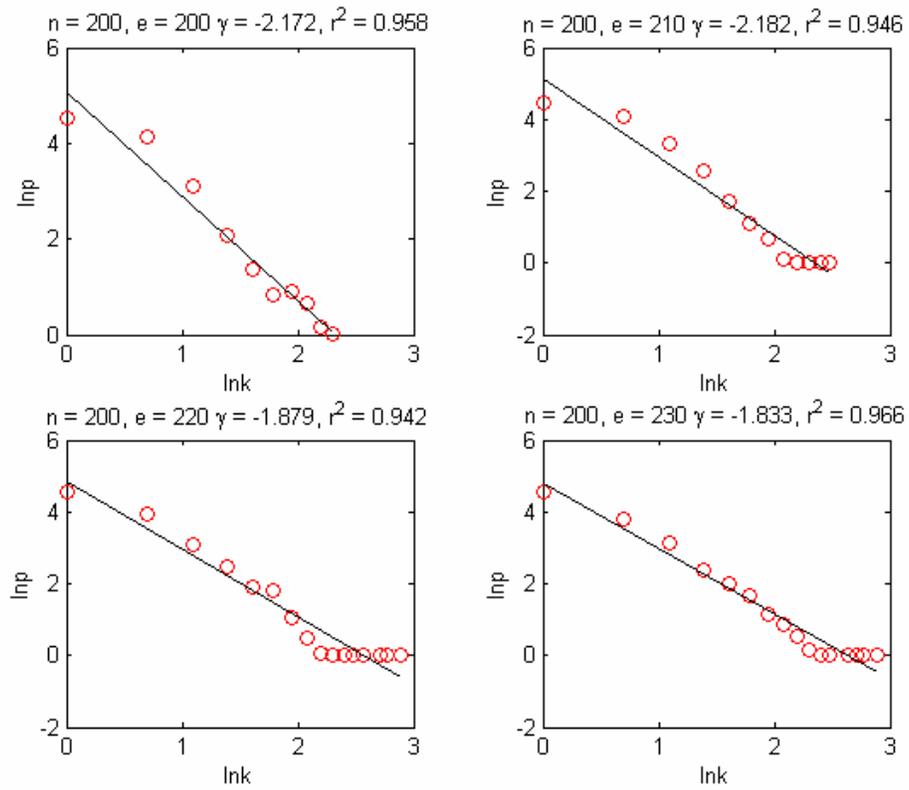

**(a)**

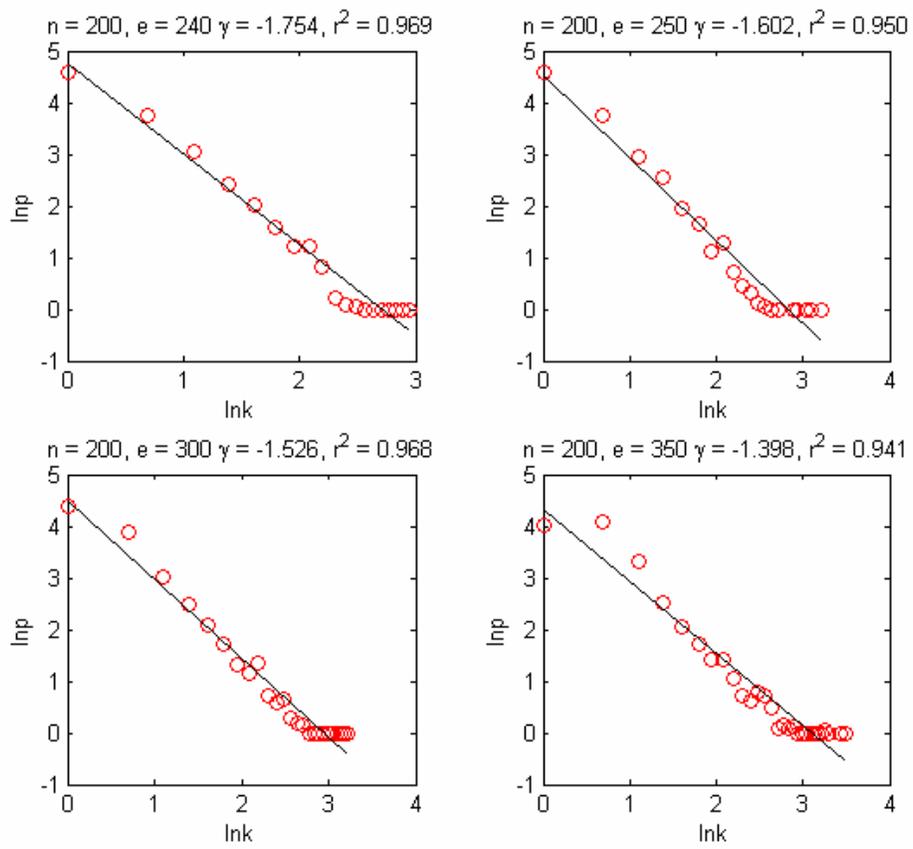

**(b)**

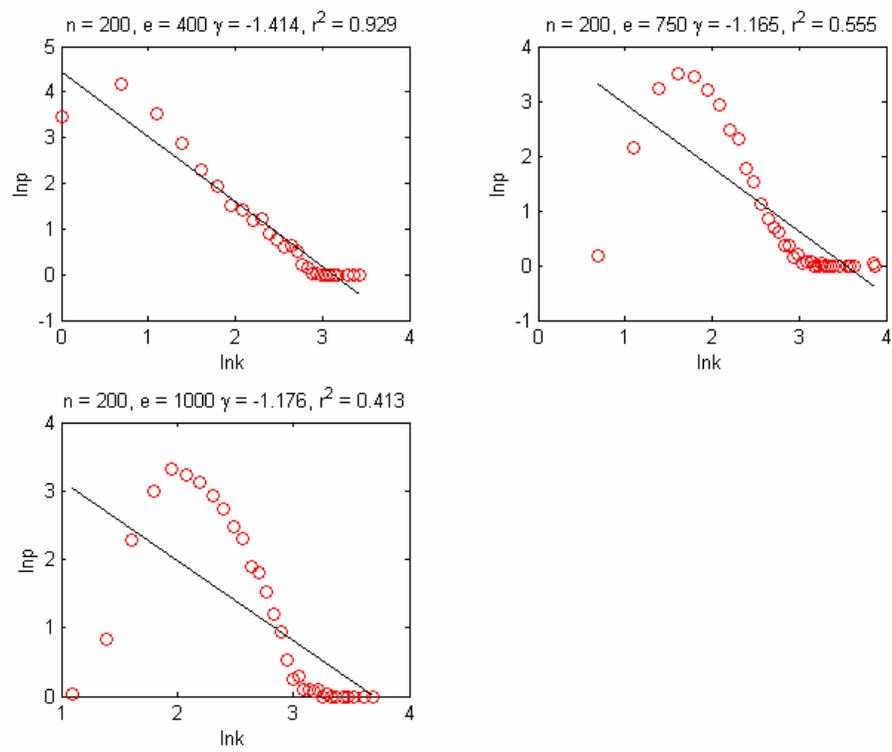

**(c)**